# Calibration and localization of optically pumped magnetometers using electromagnetic coils

Joonas Iivanainen[1,2*], Amir Borna[1], Rasmus Zetter[2], Tony R. Carter[1], Julia M. Stephen[3], Jim McKay[4], Lauri Parkkonen[2], Samu Taulu[5] and Peter D.D. Schwindt[1]

[1]Sandia National Laboratories, PO Box 5800, Albuquerque, NM 87185–1082, United States of America

[2]Aalto University School of Science, FI-00076, Aalto, Finland

[3]The Mind Research Network a Division of Lovelace Biomedical Research Institute, Albuquerque, NM 87106, United States of America

[4]Candoo Systems Inc., Port Coquitlam, BC, Canada

[5]University of Washington Seattle, Seattle, WA, United States of America

*Correspondence: jaiivan@sandia.gov

**Abstract**

In this paper, we propose a method to estimate the position, orientation and gain of a magnetic field sensor using a set of (large) electromagnetic coils. We apply the method for calibrating an array of optically pumped magnetometers (OPMs) for magnetoencephalography (MEG). We first measure the magnetic fields of the coils at multiple known positions using a well-calibrated triaxial magnetometer and model these discretely sampled fields using vector-spherical harmonics (VSH) functions. We then localize and calibrate a sensor by minimizing the sum of squared errors between the model signals and the sensor responses to the coil fields. We show that by using homogeneous and first-order gradient fields, the sensor parameters (gain, position, and orientation) can be obtained from a set of linear equations with pseudo-inverses of two matrices. By determining the coil currents based on the VSH models, the coil fields can approximate these low-order field components, yielding computationally simple initial estimates of the sensor parameters. As a first test of the method, we placed a fluxgate magnetometer at multiple positions and estimated the RMS position, orientation, and gain errors of the method to be 1.0 mm, 0.2° and 0.8%, respectively. Last, we calibrated a 48-channel OPM array. The accuracy of the OPM calibration was tested by using the OPM array to localize magnetic dipoles in a phantom, which resulted in an average position error of 3.3 mm. The results demonstrate the feasibility of using electromagnetic coils to calibrate and localize OPMs for MEG.

## 1. Introduction

Magnetoencephalography (MEG) is a functional neuroimaging technique in which the magnetic fields of electrically active neuron populations in the human brain are detected outside of the head (Hämäläinen et al. 1993). To estimate the active neural sources in the brain from the measured field patterns, an inverse problem must be solved, which involves modeling of the electric neural sources and their magnetic fields as detected by the MEG sensor array. To make the modeling and the source estimate precise, the sensor geometry must be accurately coregistered with the magnetic resonance images of the subject's head, i.e., precise knowledge of the sensor positions and orientations with respect to the subject's brain is needed. In addition, the sensor gains (conversion factors from volts to Tesla) must be known.

Traditional MEG systems utilize superconducting quantum interference devices (SQUIDs) that are housed inside a dewar in a rigid helmet-shaped configuration. To calibrate the SQUID sensors two types of methods have been used. In the first group of methods, large Helmholtz coils producing uniform magnetic fields have been employed to find the SQUID gains (e.g., de Melo et al. 2009; Li et al. 2016). While in the second group, small individual calibration coils are used to estimate the SQUID positions, orientations and gains (Chella et al. 2012; Adachi et al. 2014 and 2019). The rigid transformation between the known SQUID sensor positions and the subject's head is usually obtained by localizing small coils attached to the subject's scalp by digitizing the coil positions relative to known fiducial points before the measurement (e.g., Ahlfors and Ilmoniemi 1989; Hansen et al. 2010).

Recently, a new type of magnetometer with sensitivity suitable for MEG has become available. In contrast to SQUIDs, which detect the field at least ~2 cm from the scalp, these optically pumped magnetometers (OPMs) can measure the magnetic field within millimeters from the scalp, increasing the signal magnitude and the spatial resolution of the measurement (Boto et al. 2016; Iivanainen et al. 2017). The introduction of this new sensor type has led to the development of various coregistration methods (e.g. Zetter et al. 2019) with new requirements that must be taken into account. The possibility to obtain higher spatial resolution due to the shorter measurement distance motivates development of accurate methods that would capture the finer spatial details of the magnetic fields that the OPMs may provide in comparison to the SQUID-based MEG systems (Iivanainen et al. 2021). Moreover, as OPM positions can be optimized for each subject in a flexible manner, the coregistration method should be capable of resolving the individual sensor locations and orientations on a per-subject basis.

Most of the coregistration methods developed for OPM-MEG systems have used optical scanners to measure the sensor positions with respect to the head (Zetter et al. 2019; Hill et al. 2020; Gu et al. 2021; Cao et al. 2021). These methods have either coregistered the sensor helmet with the brain and used the helmet geometry to obtain the individual sensor positions or directly identified the individual sensor positions from the optical scan. The optical methods have some drawbacks as they can suffer from line-of-sight issues and cannot necessarily obtain precise OPM positions and orientations as these are difficult to obtain from geometrical models (they depend not only on the external geometry of the OPM, but also on the OPM laser beam shape, the vapor cell optical depth and on-sensor coil configuration, etc.). Additionally, a magnetic method has been proposed that uses a set of pre-measurement digitized small dipolar coils attached to the subject's head for finding the positions, orientations, and gains of the on-scalp MEG sensors with respect to the subject's head/brain (Pfeiffer et al. 2018; 2020).

Here, we propose a method that uses arbitrary external electromagnetic coils to estimate the position, orientation, and gain of a magnetic-field sensor. The method is based on modeling the coil fields using coefficients of vector spherical harmonic (VSH) expansions, which are obtained from *a-priori* measurements of the coil fields. We used the method to find the positions, orientations, and gains of OPMs in a 48-channel sensor array. Subsequently, we localized magnetic dipoles of a phantom using the calibrated sensor array and compared the localization results to the known locations of the dipoles.

## 2. Background

In this section, we lay out the methodology for obtaining the gain as well as the 3D position and orientation of a magnetic field sensor (from here on referred to as *sensor parameters*) using magnetic fields generated by a set of $N$ electromagnetic coils. The basic idea behind the method is straightforward. We measure the sensor response when each coil is excited. We model the magnetic fields of the coils and find the sensor parameters that minimize the sum of squared errors between the measurements and the models.

More formally, we denote the measured sensor response of the $i$'th coil as $y_i$. We model the sensor output (in volts) as

$$b_i = g\vec{B}_i(\vec{r}) \cdot \vec{n}, \tag{1}$$

where $g$ is the sensor gain (V/T), $\vec{r}$ is the sensor position, $\vec{n}$ is a unit vector representing the sensor orientation and $\vec{B}_i(\vec{r})$ is the modeled magnetic field of the $i$'th coil at $\vec{r}$. We seek the sensor parameters by minimizing the sum of squared errors between the models and the measurements:

$$\underset{g,\vec{r},\vec{n}}{\operatorname{argmin}} \sum_{i=1}^{N} \left(y_i - g\vec{B}_i(\vec{r}) \cdot \vec{n}\right)^2. \tag{2}$$

### 2.1. Obtaining magnetic field models

Accurate determination of the sensor parameters with Eq. 2 relies on accurate models of the magnetic fields. In principle, the magnetic fields could be calculated by using computational models of the coils. However, such an approach is likely inaccurate as it is difficult to correctly model factors such as coil nonidealities and magnetic field interaction with the surroundings (especially prominent when the coils are inside a magnetic shield causing polarization effects). Therefore, it is preferable to base the models on measurements of the coil magnetic fields at multiple positions. To obtain the models, the coil magnetic fields must be measured at multiple positions with a vector magnetometer whose parameters are known accurately.

As the parameter space of the optimization problem (Eq. 2) is continuous, the representation of the coil magnetic field must be continuous as well; the discrete measurements of the magnetic field should allow for an accurate interpolation of the field. For example, the field model could be obtained by a linear interpolation if the field is mapped using sufficiently dense three-axis measurements. Another option is to model the field as a linear combination of basis functions. When the basis-function coefficients are known, the field can be interpolated. Here, we opt for the latter approach and model the magnetic field using a linear combination of vector spherical harmonics (VSHs). Compared to other interpolation methods or basis-function expressions, the VSH model has some advantages. First, when the coils are sufficiently far away from the sensor, their field energy will be on the lowest VSH orders; thus, only a relatively low number of VSH components is sufficient for modeling as compared to models where such a clear hierarchical structure is not achievable. Second, because the expansion is based on the physical characteristics of magnetic fields (given by Maxwell's equations), such an expansion offers some robustness against noise.

In a source-free space, the magnetic fields of far-away sources (the magnetic field coils) can be represented with VSHs as (e.g., Hill 1954; Taulu & Kajola 2005)

$$\vec{B}_i(\vec{r}) = -\mu_0 \sum_{l=1}^{\infty} \sum_{m=-l}^{l} \beta_i^{lm} r^{l-1} \sqrt{l(2l+1)} \vec{W}_{lm}(\theta, \varphi) = \sum_{l=1}^{\infty} \sum_{m=-l}^{l} \beta_i^{lm} \vec{w}_{lm}(r, \theta, \varphi), \quad (3)$$

where $\mu_0$ is the permeability of free space, $l$ is the degree and $m \in \{-l, -l+1, \ldots, l-1, l\}$ is the order of the VSH, $\vec{W}_{lm}$ (or $\vec{w}_{lm}$), $(r, \theta, \varphi)$ are the spherical coordinates, and $\beta_i^{lm}$ are the VSH coefficients. The VSHs and their coefficients can be scaled arbitrarily; we opt for the latter expression in Eq. 3 due to its simplicity. The VSHs corresponding to $l=1$ are the three homogeneous components of the field along the Cartesian coordinate axes while those corresponding to $l=2$ are the five first-order gradients. Higher degrees correspond to higher-order gradients. The squared coefficients $\{\beta_i^{lm^2}\}$ comprise the VSH spectrum of the magnetic field.

The continuous field $\vec{B}_i(\vec{r})$ can then be represented with a set of VSH coefficients that can be computed from the measurements of the coil field at known positions as (e.g., Taulu et al. 2004)

$$\boldsymbol{\beta}_i = \mathbf{S}^\dagger \mathbf{m}_i, \quad (4)$$

where $\boldsymbol{\beta}_i$ is a $C \times 1$ vector containing $C$ VSH coefficients, $\mathbf{S}$ is an $M \times C$ matrix containing the $\vec{w}_{lm}$ patterns as detected by the $M \times 1$ measurements $\mathbf{m}_i$, and $\mathbf{S}^\dagger$ denotes the pseudo-inverse of $\mathbf{S}$.

*2.2. Simplifying the optimization problem*

Even if the obtained coil models were perfectly accurate, it is not guaranteed that a magnetic field sensor can be calibrated with the method as the optimization of Eq. 2 using the field models may not converge to the correct sensor parameters (~global optimum). Generally, the convergence will depend on the number of coils used and their VSH spectrum. Next, we will briefly describe how the optimization problem can be simplified.

If the coils generate low-order VSH components, the sensor parameters can be solved using linear equations. Specifically, we assume that $N_H$ coils generate homogeneous field components (linear combinations of VSHs corresponding to $l = 1, m = -1, 0, 1$) and $N_G$ coils generate first-order gradients ($l = 2, m = -2, -1, 0, 1, 2$) and homogeneous components. As shown in the Appendix, the sensor parameters can then be obtained as:

$$\mathbf{g} = \mathbf{H}^\dagger \mathbf{b}_H \tag{5}$$

$$\mathbf{r} = \mathbf{G}^\dagger (\mathbf{b}_G - \mathbf{H}_G \mathbf{g}), \tag{6}$$

where $\mathbf{g} = g\mathbf{n}$, the sensor orientation unit vector $\mathbf{n}$ multiplied by the sensor gain $g$, $\mathbf{r}$ is the sensor position vector, $\mathbf{H}$ is an $N_H \times 3$ matrix describing the uniform fields of the homogeneous coils, $\mathbf{G}$ is an $N_G \times 3$ matrix whose elements depend on $\mathbf{g}$ and the first-order gradients of the coils, $\mathbf{H}_G$ is an $N_G \times 3$ matrix that contains the homogeneous components of the gradient fields, and $\mathbf{b}_H$ and $\mathbf{b}_G$ are $N_H \times 1$ and $N_G \times 1$ vectors comprising the sensor responses to the homogeneous and gradient fields, respectively.

In theory, the sensor parameters should be obtainable by three homogeneous fields that are not collinear (preferably they are orthogonal) and three gradient fields that span three VSH gradient coefficients. Intuitive explanation for the case where each of the coils generates a distinct VSH component is as follows. Three perfectly homogeneous fields along *x*, *y*, and *z* encode the sensor gain and orientation as these fields do not depend on the position in space. When the sensor orientation and gain are known, the first-order linear gradients encode the sensor position. As a point of practical interest, six coils in a three-axis Helmholtz configuration would not typically give a robust localization performance. Such setup can produce the three homogeneous fields, but it can only produce two linear gradients as just two of $dB_x/dx, dB_y/dy$ and $dB_z/dz$ are independent. Thus, a coil producing a transverse gradient (e.g., $dB_y/dx$) would be needed.

Even though realistic coils do not produce perfect homogeneous or first-order gradient fields, Eqs. 5 and 6 provide a good first-order approximation for the sensor parameters in two common cases. We will next discuss these cases.

*Compensation coil set.* Coil sets that counter external magnetic interference are commonly used in bioelectromagnetic measurements with OPMs (e.g., Holmes et al. 2018, Iivanainen et al. 2019). Typically, the coils in such systems are designed to produce homogeneous and gradient fields; the VSH spectrum of the coil field is peaked at a specific low-order component. A good initial estimate for the sensor parameters ($\mathbf{g}_0, \mathbf{r}_0$) can be obtained by using Eqs. 5 and 6 with the homogeneous and first-order gradient field coils and their VSH coefficients up to $l = 2$. The full optimization problem (Eq. 2) can be then solved using the "full" field models (i.e., all VSH coefficients up to the bandlimit $l_{\max}$ given by the limitations of the *a-priori* coil field measurement) and $\mathbf{g}_0$ as well as $\mathbf{r}_0$ as initial estimates for the sensor parameters. As the low-order VSH coefficients should capture most of the energy of such coil fields, $\mathbf{g}_0$ and $\mathbf{r}_0$ should be close to the true sensor parameters. The full optimization then "fine-tunes" the sensor parameters.

*Arbitrary coil arrangement.* The first-order approximation is also useful in the case of an arbitrary coil arrangement (e.g., an array of magnetic dipoles). When the VSH coefficients of the *N* coils are known, $\mathbf{B} = [\boldsymbol{\beta}_1, ..., \boldsymbol{\beta}_N]$, the coil currents $\mathbf{i}$ that produce a field having approximately the desired distribution of the VSH coefficients $\widetilde{\boldsymbol{\beta}}$ can be obtained with the pseudo-inverse as (Simola and Taulu 2018)

$$\mathbf{i} = \mathbf{B}^\dagger \widetilde{\boldsymbol{\beta}}. \tag{7}$$

When the coil currents are set so that approximately homogeneous fields and first-order gradients are generated, $\mathbf{g}_0$ and $\mathbf{r}_0$ can be computed using Eqs. 5 and 6. Using $\mathbf{g}_0$ and $\mathbf{r}_0$ as the initial estimate, the optimization problem (Eq. 2) can be solved using the full VSH models up to $l_{\max}$ and the sensor responses to all individual *N* coils. We note that not all coil arrangements will allow one the generation of the desired low-order VSH components. Thus, it might be beneficial to calculate the condition number of matrix $\mathbf{B}$ and add coils to the arrangement if the rank is low.

*2.3. Summary of the method*

As a summary, the proposed method for estimating the sensor parameters consists of the following steps:

1. Measure the magnetic fields of the *N* coils with a well calibrated 3D sensor (e.g., fluxgate magnetometer) over the region of the sensor array (the OPMs). Fit VSH models to the measurements.
2. Using the VSH models, compute the currents for exciting homogeneous fields and first-order gradients.
3. Measure the response of the sensor (which is to be calibrated) to the homogeneous fields and gradients.
   a. Compute $\mathbf{g}_0$ and $\mathbf{r}_0$ using the lowest-order VSH coefficients with Eqs. 5 and 6.
   b. (Optional): Fine-tune the estimates $\mathbf{g}_0$ and $\mathbf{r}_0$ by optimizing Eq. 2 with full VSH models of the homogeneous and gradient fields.
4. Excite coils individually.
   a. Find the sensor parameters by optimizing Eq. 2 and using the full VSH models. Use $\mathbf{g}_0$ and $\mathbf{r}_0$ (from step 3a or 3b) as initial estimates for the sensor parameters.

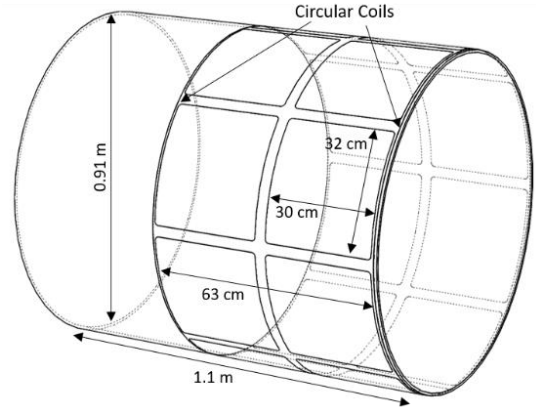

**Figure 1: Schematic illustration of the 16 rectangular and 2 circular coils embedded in the coil form within the cylindrical shield housing a six-sensor optically pumped magnetometer array. For an illustration of the OPM-MEG system together with the cylindrical shield please see Figure 2 in Borna et al. (2017).**

### 3. Methods

In this Section, we outline the methods for calibrating an array of OPM sensors residing inside a magnetic shield. The OPM system is described in detail in Borna et al. (2017, 2020); for an illustration of the system please see Figure 2 in Borna et al. (2017). Briefly, the system consists of six OPM sensors (Johnson et al. 2010; Colombo et al. 2016) each with 4 channels within a sensor-array holder that can accommodate nine sensors. Each OPM can measure a magnetic field component on the plane perpendicular to the optical axis of the sensor. With respect to the head, the field components tangential to the scalp are measured. Typically, the two orthogonal tangential components are measured separately. In total, the system has 48 channels (six sensors with 4 channels measuring two field components). The OPMs reside inside a person-sized magnetic shield with 18 embedded electromagnetic coils (see Fig. 1).

We first measure the magnetic fields of the coils within the sensor array holder using a triaxial fluxgate magnetometer. We then fit VSH models to these measurements. We validate the calibration method by localizing the fluxgate at different positions within the sensor array holder as explained in Sec. 3.3 below. Last, we calibrate the OPM array using the method and utilize the OPM array to localize magnetic dipoles within a phantom.

*3.1. Measurement of the coil fields*

We measure the magnetic fields of the 18 coils in the vicinity of the sensor-array holder using a three-axis fluxgate magnetometer (FGM3D, Sensys GmbH, Bad Saarow, Germany). The fluxgate has three channels measuring the orthogonal field components at different locations along the fluxgate axis with a spacing of 2 cm. The channels are orthogonal within ±0.5°, while their gain is 0.1 V/µT ± 0.1%.

Altogether we measure 108 locations, giving in total 324 measurements of the magnetic field of each coil. VSH coefficients up to order $l = 5$ (35 components) are estimated from the measurements using Eq. 4. The VSH functions are generated using bfieldtools Python software package (Mäkinen et al. 2020; Zetter et al. 2020). We normalize the VSHs so that each component corresponds to one unit of magnetic energy ("energy" normalization option in bfieldtools). We set the average coordinates of the measurement positions as the origin of the VSH expansion. This choice of origin was selected to get the zero crossings of the gradient fields into the sensor array volume. We use the root-mean-square (RMS) error normalized by the RMS value of the data to assess the goodness of the VSH fits.

*3.2 Calibration methodology*

We use the methodology summarized in Sec. 2.3 to estimate the parameters of the fluxgate magnetometer channels and the OPM sensors. We measure the responses of the sensors (that are to be calibrated) to the coil fields. The coils are excited sequentially using sinusoidal currents with a frequency of 20 Hz. The sensor response amplitudes are estimated from the data using lock-in detection.

Of the 18 shield coils, 17 are used in the measurements as one of the coils had a broken connector. The 17 coils are driven either individually or in "superposition" configured to generate the eight first-order VSH fields; the coil currents to generate the first eight VSH components are computed using Eq. 7 with the estimated VSH coefficients.

The initial sensor parameter estimates can be obtained with Eqs. 5 and 6 by using the data generated with the first-order VSH fields. The model parameters (the homogeneous field and gradient amplitudes) are estimated from the fitted VSH models. Subsequently, the sensor parameter estimates are "fine-tuned" by optimizing Eq. 2 with the first-order VSH field data and the full VSH models. We perform the optimization using SciPy Python library (Virtanen et al. 2020) with function *minimize* with default parameters and the Nelder–Mead algorithm. Last, the final sensor parameters are obtained by optimizing Eq. 2 with the sensor responses to the 17 individual coil fields and full VSH models of the coil fields.

*3.3 Fluxgate validation*

We position the fluxgate sensor within the three empty sensor slots of the array (six of the nine sensor holders have an OPM inserted). In total, six different positions (two different positions in one slot) are used for the fluxgate, amounting to 18 channels at different positions to calibrate. The currents of the coils are set so that the maximum field amplitude was roughly 12.5 nT over the sensor-array holder. We used the proposed method to calibrate the fluxgate channels. The known gains and relative positions/orientations of the fluxgate channels are used to estimate the calibration error of the method.

*3.4 Calibration of the OPM array*

The OPMs are configured to measure either one of the tangential components (denoted as *Bx* or *By*), and the coils are excited individually or in superposition with a maximum amplitude of 0.5 nT. Altogether four measurements corresponding to the different sensor/coil configurations (e.g., *Bx* and first-order VSH fields) are performed. The OPM data are recorded using the DAQ system described in Borna et al. (2017, 2020). The data sampling rate of the lock-in amplifier output that serves as the OPM magnetometer signal is set to 1 kHz.

The fact that the OPMs we use can measure two components of the magnetic field at the same position, motivates us to use two different approaches for the optimization. In the first approach, we optimize the two measurements ($B_x$ and $B_y$) separately while in the second approach we perform the optimization jointly for the two measurements by constraining the optimization procedure so that the "two sensors" share the same position.

We compare the estimated positions of the OPMs to a CAD model of the sensor array by aligning the obtained positions to those of the CAD model using a unit-scale rigid-body transformation obtained with the Umeyama algorithm (Umeyama et al. 1991).

*3.5 Localizing magnetic dipoles in a phantom*

We use the OPM array and the obtained sensor parameters to localize small circular coils in a phantom. The 3D-printed phantom has nine hand-wound circular coils with a diameter of 5 mm (see Fig. 6A). The phantom coils are sequentially driven with sinusoidal currents at 20 Hz, and the amplitudes are extracted from the OPM data using lock-in detection. The measurement is performed independently for $B_x$ and $B_y$; for localization the measured amplitudes are combined to give a total of 48 channels.

The coils are localized using the dipole fitting routine in the FieldTrip toolbox (function *ft_dipolefitting*; Oostenveld et al. 2011) with a grid search followed by non-linear optimization. The coils are modeled as ideal point-like magnetic dipoles. The OPM channels that have latency-adjusted time-domain correlation lower than 0.9 with the reference signal fed to the dipole are discarded from the dipole-fit procedure. These channels are deemed unreliable because of the phase offset likely caused by the OPM cross-axis projection error (Borna et al. 2021).

Last, we compare the estimated positions of the magnetic dipoles to a CAD model of the phantom. To compare the relative positions of those, the estimated positions are aligned with the CAD model using the Umeyama algorithm (Umeyama et al. 1991).

## 4. Results

*4.1. Modeling the magnetic field measurements with VSHs*

Figure 2 shows the measured magnetic field of one of the shield coils as well as its interpolation using the VSH model. The magnetic field points approximately to the same direction in the measurement volume with a gradient along the *z*-axis. The normalized RMS error of the VSH reconstruction is 1.0%. The VSH spectrum shown in the figure reveals that the dominant field components are at low orders. The homogeneous components corresponding to $l = 1$ have the highest energy followed by the first-order gradients ($l = 2$). The VSH spectrum averaged over the coils (Fig. 2C) confirms that the coils, on average, produce mostly homogeneous and first-order gradient fields. The normalized RMS error is less than 2% for all coils indicating good fits of the VSH models to the measurements.

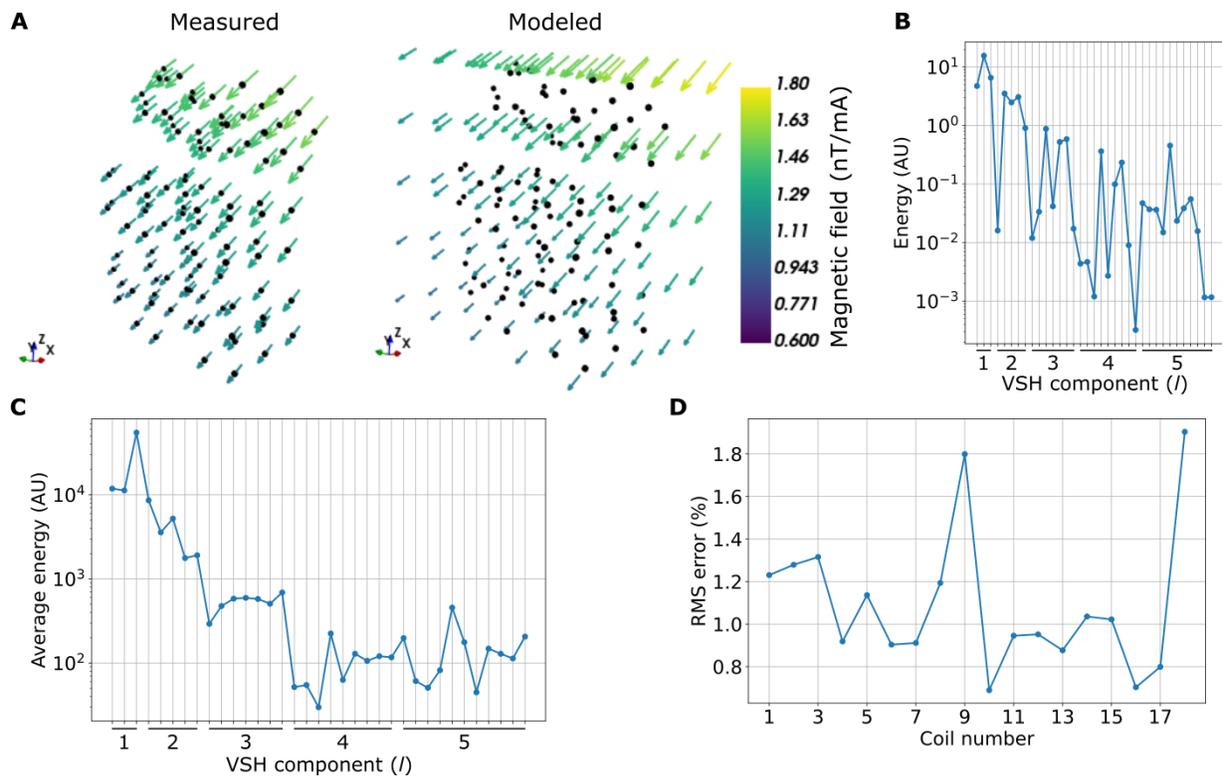

Figure 2: Fitting vector spherical harmonics (VSHs) to the measured magnetic field data. A: Example of a coil magnetic field measured with a three-axis fluxgate at 108 points together with its VSH fit extrapolated to a volume surrounding the measurement points. B: Magnetic field spectrum of the coil field in panel A given by the squared VSH coefficients. The squared coefficients are shown for each degree *l* and its corresponding *2l+1* orders. C: Average magnetic field spectrum of the 18 shield coils. D: Normalized RMS errors between the VSH fits and the measurements.

Figure 3 shows examples of calculated coil currents that approximately generate specific low-order VSH components at the measurement volume. VSH spectra of the fields peak at specific components, confirming that the currents mainly produce the desired components.

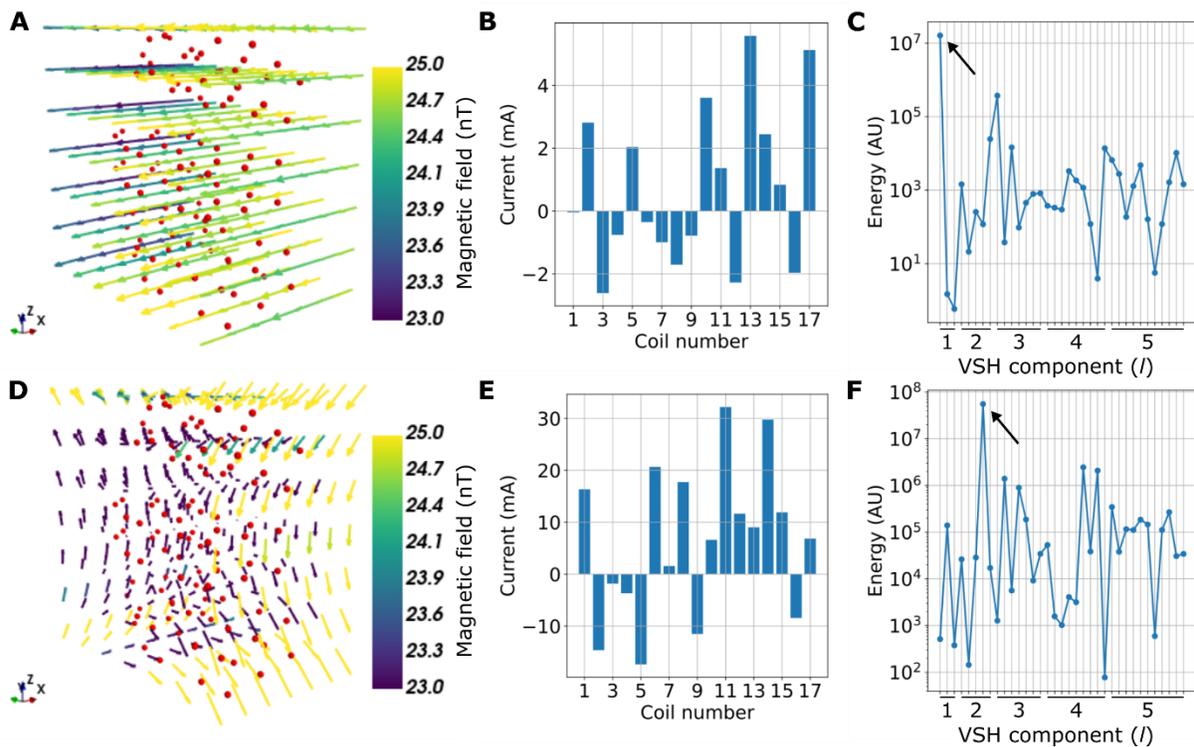

Figure 3: Examples of calculated coil currents of the 17 shield coils that generate specific vector spherical harmonic (VSH) field components. A: Example of a homogeneous magnetic field in the vicinity of the measurement points shown as red dots. B: The coil currents that generate the field shown in A. C: The VSH spectra of the generated field. The black arrow indicates the energy of the desired VSH component. D–F: Same as A–C but for a first-order gradient field.

*4.2. Fluxgate validation*

Figure 4 shows the estimated fluxgate position, orientation, and gain errors across the 6 measurement positions (18 channels). Clearly, the errors become smaller as more iterations are performed and as more VSH coefficients are taken into account in the optimization. For the final estimates, the average gain, orientation, and position errors are 0.8% (0.1%–1.1%), 0.1° (0.01°–0.4°) and 0.8 mm (0.1–2.0 mm), respectively, while the RMS errors are 0.8%, 0.2° and 1.0 mm.

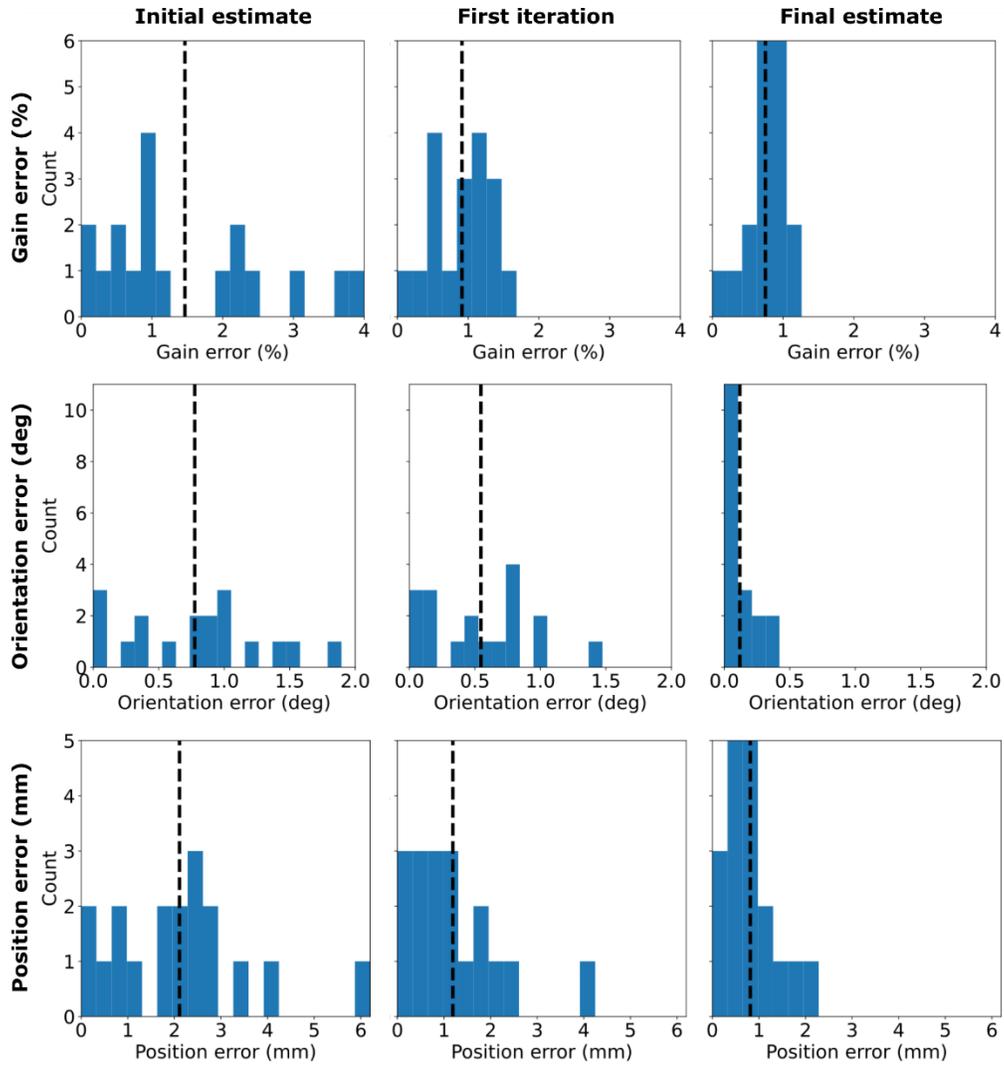

**Figure 4:** Estimated gain (top row), orientation (middle) and position (bottom) errors for the three channels of the fluxgate magnetometer at six different placements of the fluxgate. Each column of the figure shows the error histograms after different iterations. The initial estimate is given by the matrix pseudo-inversions using the first-order VSH measurements and models. The first-iteration estimate is obtained after fine tuning the initial estimate to account for the full VSH spectra of the approximate first-order VSH fields. The final estimate is obtained after optimizing the first-iteration estimate with all individual coil fields and their full VSH spectra. The vertical dashed lines indicate the average errors.

*4.3. Calibration of the OPM array and phantom localization*

The optimized OPM sensor parameters when the OPM channels sensing different field components are either treated jointly or separately are shown in Fig. 5A–B. The sensor positions visually resemble those we expect based on the CAD model of the sensor array while the obtained orientations are not orthogonal as also suggested by the characteristics of the magnetic fields of the on-sensor coils (see, e.g., Borna et al. 2017). Interestingly, when the channels sensing *Bx* and *By* are treated separately they do not localize to the exact same positions. The maximum differences between the parameters obtained jointly or separately are 0.7%, 0.3° and 4.2 mm for gain, orientation, and position, respectively.

Fig 5C compares the jointly obtained sensor positions to those of the CAD model. Altogether, the positions look rather similar, however, the calibrated positions are somewhat more scattered on the plane defined by the sensor-wise channel positions. The calibrated positions for the four channels corresponding to one sensor sit on a single plane quite well. The average distance between the calibrated and CAD positions is 2.8 mm while the range is 0.9–5.4 mm.

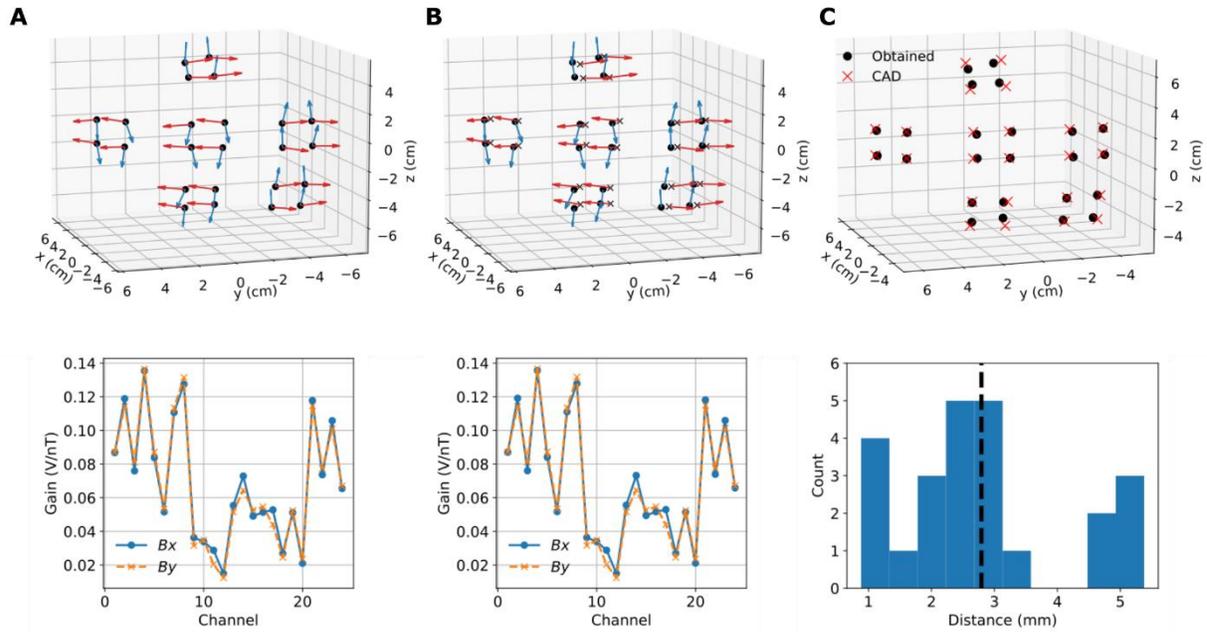

Figure 5: Obtained OPM sensor parameters. A: Sensor parameters (top: position, orientation; bottom: gain) obtained by joint optimization, i.e., assuming that the channels sensing either of the tangential field components ($B_x$ or $B_y$) share the same location. B: Sensor parameters obtained by treating the channels separately, i.e., $B_x$ and $B_y$ channels have independent location. The channel positions for $B_x$ or $B_y$ are shown as black dots and crosses, respectively. C: Comparison of the sensor positions in panel A to the CAD model of the sensor array. Histogram illustrates the distances between the positions; vertical dashed line shows the average distance.

The jointly obtained OPM parameters are used to localize magnetic dipoles in a phantom. Figure 6A shows a drawing of the phantom illustrating the locations of the nine magnetic dipoles. Figure 6B shows the estimated positions and moments of the magnetic dipoles. Visually, the estimated positions correspond well to those of the CAD model and the estimated orientations of the coil magnetic moments agree with the geometry of the phantom. The average error between the estimated dipole positions and those of the CAD model is 3.3 mm while the range is 1.1–5.7 mm (Fig. 6C).

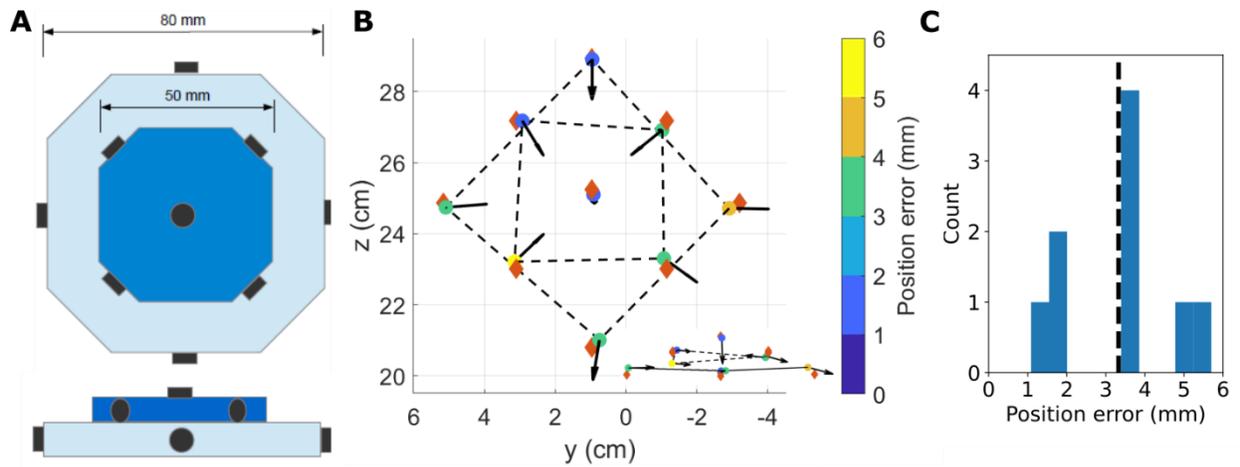

Figure 6: Localization of magnetic dipoles in a phantom using the calibrated 48-channel OPM array. A: A drawing of the 3D-printed phantom showing the locations and orientations of the small circular coils approximating magnetic dipoles (shown black). Both top and side views of the phantom are shown. B: The estimated dipole positions (circles) and orientations of the coil dipole moments (black arrows). Red diamonds show the positions of the coils according to the CAD model of the phantom. The circles are color-coded according to the distance between the estimated position and that of the CAD model. Inset gives a side view of the dipoles. C: Histogram of the position errors between the CAD model and the estimates. Dashed vertical line shows the average error.

## 5. Discussion

We proposed a method that uses magnetic fields to find the gain/calibration, position and orientation of a magnetic field sensor. In particular, we proposed a means to estimate the sensor parameters by using homogeneous and first-order gradient fields (VSHs up to $l = 2$), which can be generated with specific electromagnetic coils or by driving an appropriate arrangement of multiple individual magnetic field generators. Importantly, if only low-order VSHs are modeled, the sensor parameters can be obtained by simple matrix pseudo-inversions. The sensor-parameter estimates can then be enhanced by a more involved optimization taking into account the higher-order spatial structure of the generated magnetic fields. The method was able to resolve relative positions and orientations of fluxgate magnetometer channels with RMS errors of 1.0 mm and 0.2°, respectively, while the gain was estimated with an RMS error of 0.8%. The method yielded OPM parameters that were close to our expectations; the OPM parameters allowed us to localize magnetic dipoles with an average error of 3.3 mm.

We note that we lack the ground truth against which to compare the OPM sensor parameters. For example, we expect that the CAD positions do not perfectly reflect the real channel positions. The CAD model uses the center of the laser beam profile in the vapor cell as the channel position. However, the 'magnetic center-of-mass' of the channel will be defined by the spin polarization profile in the cell. It is our understanding that the spin polarization is not homogeneous in the vapor cell and is larger on the side of the cell where the pump laser beam enters the cell, shifting the center-of-mass of the channel towards the cell wall. Thus, computing the distances between the CAD model and the obtained positions shown in Fig. 6C probably overestimates the position errors.

Using the calibrated 48-channel OPM array, we were able to localize 5-mm diameter hand-wound circular coils (magnetic dipoles) in a custom phantom with an average error of 3.3 mm. For comparison, a commercial MEG system comprising 102 SQUID magnetometers and 204 planar gradiometers (MEGIN Oy, Helsinki, Finland) has been shown to localize current dipoles in a commercial phantom (MEGIN Oy) with a maximum error of 2 mm (Helle et al. 2021) or with an average error between 1.6–3.9 mm (Murthy et al. 2014; depending on the co-registration methodology different average errors were achieved). Using the same phantom and optical coregistration of nine OPM sensors/channels (QuSpin Inc., Louisville, CO, USA), Zetter et al. 2019 were able to localize current dipoles with an average error of 2.1 mm. More recently, Boto et al. 2022 combined non-synchronous tri-axial OPM measurements (QuSpin Inc.; 25 sensors; 75 channels) to localize current dipole in a custom phantom with an error of 5.2 mm. Our localization error is thus similar to what has been achieved in the previous studies though our result is not directly comparable to them due to differences in the study methods (number of channels, used phantom). Thus it is of interest to test the method with similar methods as used previously and further validate it for MEG.

Ultimately, the OPM response nonlinearities, such as the cross-axis projection error (Borna et al. 2021), may limit the accuracy at which the OPM sensor parameters can be obtained with methods based on magnetic fields. However, as shown here, these errors may be alleviated to some degree by constraining the optimization by combining multi-axis measurements. The sensor parameters obtained by using combined channel positions showed smaller average dipole localization error than those obtained with separate channel positions (3.3 vs 4.0 mm; result not shown here). Additionally, the bit resolution of the digital-to-analog converters driving the coils may have limited how accurately we can implement the currents to generate the low-order VSH components (Fig. 3) for the OPM sensors, reducing the localization accuracy compared to the fluxgate magnetometer.

The requirements for sensor parameter accuracies in (on-scalp) MEG are still not fully understood. Generally, it can be stated that the required accuracy will depend on the signal-to-noise ratio and the desired spatial bandwidth (or spatial cutoff frequency) (e.g., Ahonen et al. 1993; Iivanainen et al. 2021) as higher spatial frequencies (or "spatial field modes") will be more sensitive to errors. A recent simulation study suggested that to obtain similar or higher accuracy with on-scalp MEG than with SQUID-MEG, RMS sensor position and orientation errors should be <4 mm and <10°, respectively (Zetter et al. 2018). The results obtained here suggest that the proposed method can fulfill these requirements.

With the low-order field models, the sensor parameters can be obtained by pseudo-inversion of two low-dimensional matrices. Due to the small number and low computational cost of the operations needed, the method might be well suited for fast real-time tracking of the sensor parameters. Such tracking may be accomplished by feeding sinusoids at different frequencies to the coils with subsequent lock-in detection of the coil amplitudes. For the approach, a frequency band must be reserved for the coil signals; the required bandwidth will depend on the number of coils used and the desired time resolution. For the tracking approaches, it may be beneficial to first run the full optimization routine proposed here using static sensors to determine the sensor parameters as accurately as possible. For real-time tracking, the sensor parameters can then be updated with the first-order model. The model could also incorporate, for example, some priors based on how fast the sensor parameters are likely to change.

We used a comprehensive fluxgate measurement of the coil magnetic fields to determine their VSH field models. We note that the fluxgate measurement of the coil fields was performed in 2019 while the validation with the fluxgate and calibration of the OPM array were done in 2021. It seems that once measured, the coil VSH models stay quite accurate even though the shield is opened and degaussed multiple times in that period. There are three compelling alternatives to the comprehensive but rather cumbersome fluxgate measurement. First, the fluxgate measurement positions could be optimized to yield the desired VSH coefficients with as low a number of measurements as possible. Second, with optical tracking of the fluxgate position, a large number of measurements of the coil fields could be collected rather quickly (see Mellor et al. 2021 and Rea et al. 2021). The spatial oversampling of the coil fields could reduce the effects of random fluxgate positioning errors on the obtained VSH coefficients. Last, instead of treating the sensors (to be calibrated with the method) independently, the whole array could be treated jointly to leverage the spatial sampling provided by the array to directly estimate the VSH coefficients and sensor parameters from the excited coil fields without a separate *a priori* measurement of the coil fields. This approach would probably benefit from a large number of sensors in the array.

**Conclusions**

We propose a method that uses external magnetic field generators to localize a magnetic field sensor. The magnetic fields of the generators are modelled using vector-spherical harmonics based on *a-priori* measurements of the fields. We propose a way to estimate the gain, position, and orientation of the sensor by using homogeneous and first-order gradient fields. We further show how these estimates can be enhanced with nonlinear optimization. The method allows us to estimate the relative positions, orientations, and gains of the channels of a fluxgate magnetometer with RMS errors of 1.0 mm, 0.2° and 0.8%, respectively. We use the method to calibrate a 48-channel optically pumped magnetometer array; the calibrated array allows us to localize magnetic dipoles with an average error of 3.3 mm.

**Funding information**

This work has received funding from the National Institute of Biomedical Imaging and Bioengineering of the National Institutes of Health under award number U01EB028656 and the European Research Council under grant agreement No 678578 (project HRMEG).

**Author contributions**

Conceptualization, JI, AB, RZ, JM, LP, ST, PDDS; methodology, JI; software, JI, AB, RZ; validation, JI; formal analysis, JI; investigation, JI, AB; resources, JI, AB, TRC, JM, PDDS; data curation, JI, AB; writing—original draft preparation, JI; writing—review and editing, JI, AB, RZ, JMS, JM, LP, ST, PDDS; visualization, JI; supervision, LP, ST, PDDS; project administration, LP, ST, PDDS; funding acquisition, AB, JMS, LP, ST, PDDS. All authors have read and agreed to the published version of the manuscript.

**Conflict of interest**

Authors declare no conflict of interest.

**Data availability statement**

The scripts for sensor array calibration are available from the corresponding author on reasonable request.

**Acknowledgements**

Sandia National Laboratories is a multimission laboratory managed and operated by National Technology & Engineering Solutions of Sandia, LLC, a wholly owned subsidiary of Honeywell International Inc., for the U.S. Department of Energy National Nuclear Security Administration under contract DENA0003525. This paper describes objective technical results and analysis. Any subjective views or opinions that might be expressed in the paper do not necessarily represent the views of the U.S. Department of Energy, the United States Government, the National Institutes of Health, or European Research Council.

**Appendix**

Here, we derive Eqs. 5 and 6 for obtaining the sensor parameters with the first-order VSH models. The vector sensor gain is $\vec{g} = g\hat{n} = g_x\hat{e}_x + g_y\hat{e}_y + g_z\hat{e}_z$, where hat denotes unit vector and $\hat{e}_x$ denotes the unit vector along the x-axis. Suppose that we have a homogeneous field $\vec{B}_i = H_{i,x}\hat{e}_x + H_{i,y}\hat{e}_y + H_{i,z}\hat{e}_z$, now the sensor response to the field is

$$b_i = \vec{g} \cdot \vec{B}_i = H_{i,x}g_x + H_{i,y}g_y + H_{i,z}g_z. \tag{A.1}$$

If we have a collection of these equations, they can be written in the matrix form as

$$\mathbf{b}_\mathrm{H} = \mathbf{H}\mathbf{g}, \tag{A.2}$$

where the elements of the vectors and matrices are defined as $\mathbf{b}_\mathrm{H}[i] = b_i$, $\mathbf{H}[i,j] = H_{i,j}$ and $\mathbf{g}[j] = g_j$ ($j \in \{x, y, z\}$ where the index mapping is as $\{x: 1, \ y: 2, \ z: 3\}$. The sensor gain can then be solved from Eq. A.2 using the pseudo-inverse as shown in Eq. 5. Please note that we use the pseudo-inverse as the matrix $\mathbf{H}$ is not necessarily square.

Next, we model the field of the *i*th gradient coil in a matrix form as

$$\begin{pmatrix} B_{x,i} \\ B_{y,i} \\ B_{z,i} \end{pmatrix} = \begin{pmatrix} G_{xx,i} & G_{yx,i} & G_{zx,i} \\ G_{yx,i} & G_{yy,i} & G_{yz,i} \\ G_{zx,i} & G_{yz,i} & -G_{xx,i} - G_{yy,i} \end{pmatrix} \begin{pmatrix} x \\ y \\ z \end{pmatrix} + \begin{pmatrix} B^H_{x,i} \\ B^H_{y,i} \\ B^H_{z,i} \end{pmatrix} \Leftrightarrow \mathbf{B}_i = \mathbf{G}_i \mathbf{r} + \mathbf{B}^H_i, \tag{A.3}$$

where the elements of the gradient matrix $\mathbf{G}_i$ are defined as $G_{xy,i} = dB_{x,i}/dy$ and $B^H_{x,i}$ are the homogeneous components of the gradient field. The gradient matrix is defined so that its elements satisfy Maxwell's equations in a source-free space $\nabla \cdot \vec{B} = 0$ and $\nabla \times \vec{B} = 0$, i.e., sum of its diagonal elements must be zero ($G_{xx,i} + G_{yy,i} - G_{xx,i} - G_{xx,i} = 0$) and it has to be symmetric ($G_{xy,i} = G_{yx,i}$). In other words, two of the three diagonal gradients are independent while three off-diagonal gradients are independent. A measurement of the *i*th gradient coil can be written as

$$b_i = \vec{g} \cdot \vec{B}_i = (\mathbf{G}_i[1,:]\mathbf{g})x + (\mathbf{G}_i[2,:]\mathbf{g})y + (\mathbf{G}_i[3,:]\mathbf{g})z + (\mathbf{B}^H_i)^T \mathbf{g}, \tag{A.4}$$

where $\mathbf{G}_i[j,:]$ denotes the *j*th row of the gradient matrix of the *i*th coil. A collection of the gradient coil measurements can then be written as

$$\mathbf{b}_\mathrm{G} = \mathbf{G}\mathbf{r} + \mathbf{H}_\mathrm{G}\mathbf{g}, \tag{A.5}$$

where the matrix elements are defined as $\mathbf{b}_\mathrm{G}[i] = b_i$, $\mathbf{G}[i,j] = \mathbf{G}_i[j,:]\mathbf{g}$ and $\mathbf{H}_\mathrm{G}[i,:] = (\mathbf{B}^H_i)^T$. Equation A.5 can then be solved to yield Eq. 6.

**References**

Adachi, Y., Higuchi, M., Oyama, D., Haruta, Y., Kawabata, S., & Uehara, G. (2014). Calibration for a multichannel magnetic sensor array of a magnetospinography system. IEEE Transactions on Magnetics, 50(11), 1-4.

Adachi, Y., Oyama, D., Terazono, Y., Hayashi, T., Shibuya, T., & Kawabata, S. (2019). Calibration of room temperature magnetic sensor array for biomagnetic measurement. IEEE Transactions on Magnetics, 55(7), 1-6.

Ahlfors, S., & Ilmoniemi, R. J. (1989). Magnetometer position indicator for multichannel MEG. *In Advances in biomagnetism* (pp. 693-696). Springer, Boston, MA.

Ahonen, A.I., Hämäläinen, M.S., Ilmoniemi, R.J., Kajola, J. E., Simola, J. T., & Vilkman, V. A . (1993). Sampling theory for neuromagnetic detector arrays. *IEEE Transactions on Biomedical Engineering*, *40*(0), 859.

Borna, A., Carter, T. R., Goldberg, J. D., Colombo, A. P., Jau, Y. Y., Berry, C., ... & Schwindt, P. D. (2017). A 20-channel magnetoencephalography system based on optically pumped magnetometers. *Physics in Medicine & Biology*, *62*(23), 8909.

Borna, A., Carter, T. R., Colombo, A. P., Jau, Y. Y., McKay, J., Weisend, M., ... & Schwindt, P. D. (2020). Non-invasive functional-brain-imaging with an OPM-based magnetoencephalography system. *PLOS ONE*, *15*(1), e0227684.

Borna, A., Iivanainen, J., Carter, T. R., McKay, J., Taulu, S., Stephen, J., & Schwindt, P. D. (2021). Cross-Axis Projection Error in Optically Pumped Magnetometers and its Implication for Magnetoencephalography Systems. *NeuroImage*, 118818.

Boto, E., Bowtell, R., Krüger, P., Fromhold, T. M., Morris, P. G., Meyer, S. S., ... & Brookes, M. J. (2016). On the potential of a new generation of magnetometers for MEG: a beamformer simulation study. *PLOS ONE*, 11(8), e0157655.

Boto, E., Shah, V., Hill, R. M., Rhodes, N., Osborne, J., Doyle, C., ... & Brookes, M. J. (2022). Triaxial detection of the neuromagnetic field using optically-pumped magnetometry: feasibility and application in children. NeuroImage, 119027.

Cao, F., An, N., Xu, W., Wang, W., Yang, Y., Xiang, M., ... & Ning, X. (2021). Co-registration Comparison of On-Scalp Magnetoencephalography and Magnetic Resonance Imaging. *Frontiers in Neuroscience*, 15.

Chella, F., Zappasodi, F., Marzetti, L., Della Penna, S., & Pizzella, V. (2012). Calibration of a multichannel MEG system based on the Signal Space Separation method. Physics in Medicine & Biology, 57(15), 4855.

Colombo, A. P., Carter, T. R., Borna, A., Jau, Y. Y., Johnson, C. N., Dagel, A. L., & Schwindt, P. D. (2016). Four-channel optically pumped atomic magnetometer for magnetoencephalography. *Optics express*, *24*(14), 15403-15416.

Gu, W., Ru, X., Li, D., He, K., Cui, Y., Sheng, J., & Gao, J. H. (2021). Automatic coregistration of MRI and on-scalp MEG. *Journal of Neuroscience Methods*, 358, 109181.

Hansen, P., Kringelbach, M., & Salmelin, R. (Eds.). (2010). *MEG: an introduction to methods*. Oxford university press.

Helle, L., Nenonen, J., Larson, E., Simola, J., Parkkonen, L., & Taulu, S. (2020). Extended Signal-Space Separation method for improved interference suppression in MEG. IEEE Transactions on Biomedical Engineering, 68(7), 2211-2221.

Hill, E. L. (1954). The theory of vector spherical harmonics. *American Journal of Physics*, *22*(4), 211-214.

Hill, R. M., Boto, E., Rea, M., Holmes, N., Leggett, J., Coles, L. A., ... & Brookes, M. J. (2020). Multi-channel whole-head OPM-MEG: Helmet design and a comparison with a conventional system. *NeuroImage*, 219, 116995.

Holmes, N., Leggett, J., Boto, E., Roberts, G., Hill, R. M., Tierney, T. M., ... & Bowtell, R. (2018). A bi-planar coil system for nulling background magnetic fields in scalp mounted magnetoencephalography. *NeuroImage*, *181*, 760-774.

Hämäläinen, M., Hari, R., Ilmoniemi, R. J., Knuutila, J., & Lounasmaa, O. V. (1993). Magnetoencephalography—theory, instrumentation, and applications to noninvasive studies of the working human brain. *Reviews of modern Physics*, 65(2), 413.

Iivanainen, J., Stenroos, M., & Parkkonen, L. (2017). Measuring MEG closer to the brain: Performance of on-scalp sensor arrays. *NeuroImage*, 147, 542-553.

Iivanainen, J., Zetter, R., Grön, M., Hakkarainen, K., & Parkkonen, L. (2019). On-scalp MEG system utilizing an actively shielded array of optically-pumped magnetometers. *NeuroImage*, *194*, 244-258.

Iivanainen, J., Mäkinen, A. J., Zetter, R., Stenroos, M., Ilmoniemi, R. J., & Parkkonen, L. (2021). Spatial sampling of MEG and EEG based on generalized spatial-frequency analysis and optimal design. *NeuroImage*, 118747.

Johnson, C., Schwindt, P. D., & Weisend, M. (2010). Magnetoencephalography with a two-color pump-probe, fiber-coupled atomic magnetometer. Applied Physics Letters, 97(24), 243703.

Li, H., Zhang, S. L., Zhang, C. X., Kong, X. Y., & Xie, X. M. (2016). An efficient calibration method for SQUID measurement system using three orthogonal Helmholtz coils. Chinese Physics B, 25(6), 068501.

Mellor, S. J., Tierney, T. M., O'Neill, G. C., Alexander, N., Seymour, R. A., Holmes, N., ... & Barnes, G. R. (2021). Magnetic Field Mapping and Correction for Moving OP-MEG. *IEEE Transactions on Biomedical Engineering*, doi: 10.1109/TBME.2021.3100770.

de Melo, C. F., Araújo, R. L., Ardjomand, L. M., Quoirin, N. S. R., Ikeda, M., & Costa, A. A. (2009). Calibration of low frequency magnetic field meters using a Helmholtz coil. Measurement, 42(9), 1330-1334.

Vema Krishna Murthy, S., MacLellan, M., Beyea, S., & Bardouille, T. (2014). Faster and improved 3-D head digitization in MEG using Kinect. Frontiers in Neuroscience, 8, 326.

Mäkinen, A. J., Zetter, R., Iivanainen, J., Zevenhoven, K. C., Parkkonen, L., & Ilmoniemi, R. J. (2020). Magnetic-field modeling with surface currents. Part I. Physical and computational principles of bfieldtools. *Journal of Applied Physics*, *128*(6), 063906.

Oostenveld, R., Fries, P., Maris, E., & Schoffelen, J. M. (2011). FieldTrip: open source software for advanced analysis of MEG, EEG, and invasive electrophysiological data. *Computational Intelligence and Neuroscience*, *2011*.

Pfeiffer, C., Andersen, L. M., Lundqvist, D., Hämäläinen, M., Schneiderman, J. F., & Oostenveld, R. (2018). Localizing on-scalp MEG sensors using an array of magnetic dipole coils. *PLOS ONE*, 13(5), e0191111.

Pfeiffer, C., Ruffieux, S., Andersen, L. M., Kalabukhov, A., Winkler, D., Oostenveld, R., ... & Schneiderman, J. F. (2020). On-scalp MEG sensor localization using magnetic dipole-like coils: A method for highly accurate co-registration. *NeuroImage*, 212, 116686.

Rea, M., Holmes, N., Hill, R. M., Boto, E., Leggett, J., Edwards, L. J., ... & Brookes, M. J. (2021). Precision magnetic field modelling and control for wearable magnetoencephalography. *NeuroImage*, 118401.

Simola, J., & Taulu, S. (2018). *U.S. Patent No. 9,977,764*. Washington, DC: U.S. Patent and Trademark Office.

Taulu, S., Kajola, M., & Simola, J. (2004). Suppression of interference and artifacts by the signal space separation method. *Brain topography*, *16*(4), 269-275.

Taulu, S., & Kajola, M. (2005). Presentation of electromagnetic multichannel data: the signal space separation method. *Journal of Applied Physics*, *97*(12), 124905.

Umeyama, S. (1991). Least-squares estimation of transformation parameters between two point patterns. *IEEE Transactions on Pattern Analysis & Machine Intelligence*, *13*(04), 376-380.

Uusitalo, M. A., & Ilmoniemi, R. J. (1997). Signal-space projection method for separating MEG or EEG into components. *Medical and Biological Engineering and Computing*, *35*(2), 135-140.

Virtanen, P., Gommers, R., Oliphant, T. E., Haberland, M., Reddy, T., Cournapeau, D., ... & Van Mulbregt, P. (2020). SciPy 1.0: fundamental algorithms for scientific computing in Python. *Nature methods*, *17*(3), 261-272.

Zetter, R., Iivanainen, J., Stenroos, M., & Parkkonen, L. (2018). Requirements for coregistration accuracy in on-scalp MEG. *Brain topography*, *31*(6), 931-948.

Zetter, R., Iivanainen, J., & Parkkonen, L. (2019). Optical Co-registration of MRI and On-scalp MEG. *Scientific Reports*, 9(1), 1-9.

Zetter, R., J. Mäkinen, A., Iivanainen, J., Zevenhoven, K. C., Ilmoniemi, R. J., & Parkkonen, L. (2020). Magnetic field modeling with surface currents. Part II. Implementation and usage of bfieldtools. *Journal of Applied Physics*, *128*(6), 063905.